\documentclass[runningheads]{llncs}
\usepackage[T1]{fontenc}
\usepackage{graphicx}
\usepackage{color}
\usepackage{listings}
\usepackage{amsmath}
\usepackage{amssymb}
\usepackage{hyperref}

\begin{document}
\title{Privacy-preserving and reward-based mechanisms of proof of engagement}
\author{Matteo Marco Montanari \and Alessandro Aldini} % \orcidID{2222--3333-4444-5555}}
%
%\authorrunning{}
\institute{University of Urbino Carlo Bo, Italy
\email{m.montanari41@campus.uniurb.it,alessandro.aldini@uniurb.it}}
\maketitle              % typeset the header of the contribution
\begin{abstract}

Proof-of-Attendance (PoA) mechanisms are typically employed to demonstrate a specific user's participation in an event, whether virtual or in-person. The goal of this study is to extend such mechanisms to broader contexts where the user wishes to digitally demonstrate her involvement in a specific activity (Proof-of-Engagement, PoE).
This work explores different solutions, including DLTs as well as established technologies based on centralized systems. The main aspects we consider include the level of privacy guaranteed to users, the scope of PoA/PoE (both temporal and spatial), the transferability of the proof, and the integration with incentive mechanisms.

%\cite{CTE_D311a_PoA}.

\end{abstract}
\section{Introduction}

The increasing interest in digitizing human processes and interactions has led to the development of new ways to certify the involvement of users in events, in both virtual and physical contexts~\cite{Marinescu2025,Joseph2021,Klopfenstein2019,Shields2017,Jovanovic2014}. Digital certificates of attendance are commonly provided in academic conferences, cultural events, and business training courses. These digital proofs of attendance (PoA) serve as verifiable, reliable, and potentially tamper-resistant evidence that a user has participated in the event. They can be used to grant access to exclusive content or improve traceability and efficiency in human resource management. 

The implementation of these mechanisms can be based on innovative technologies, such as Distributed Ledger Technologies (DLT)~\cite{Dongre2021,Chowdhury2019,ElIoini2018} and Non-Fungible Tokens (NFT)~\cite{DBLP:journals/corr/abs-2105-07447}, or traditional centralized architectures. In any case, key features to consider include the level of privacy offered to users, the possible transferability of the evidence, its spatial and temporal validity, and the integration with rewarding and incentive mechanisms.

Extending the concept of PoA, in this paper, we introduce the Proof of Engagement (PoE), the purpose of which is to digitally certify a user's involvement in any activity. PoA and, more in general, PoE find application in a variety of scenarios, from educational systems to gamification platforms, from social inclusion initiatives to more complex contexts related to the digital economy~\cite{Paspallis2018,10.1007/978-3-319-07293-7_39,Pinter2020,Klopfenstein2019}. 
%Several different scenarios fall within the scope of PoE. A user participating in a webinar might get a badge as proof of skills acquired, while loyal customers involved in a gamification context, as well as citizens performing actions that promote collective well-being and social cohesion, might be rewarded with a digital token for their activities~\cite{Pinter2020,Klopfenstein2019}. 
%Badges and tokens, while lacking direct monetary value, could be recognized by third parties, e.g., in the form of discounts, opportunities, and exclusive benefits, thus creating an ecosystem of value based on active user engagement. In particular, an important aspect that we want to emphasize and consider in this paper is the incentivizing role that the PoE might play in different situations.

\begin{example}\label{ex:intro}
As motivating examples, consider the two following scenarios.
Firstly, user Alice enrolls in a Master's course in Digital Innovation for Cultural Heritage at the Alan Smithee Institute of Higher Education. In addition to issuing a paper certificate, the Institute generates a digital badge detailing the skills Alice has acquired, which she attaches to her CV to increase her chances of accessing job opportunities. At the same time, the tourism board in Alice's country decides to offer all graduates of any higher education institute a one-time voucher, which can be used for privileged admission to exhibitions and events.
Secondly, during a vacation in Smallville, user Bob is invited by the local tourism office to evaluate the tourism services being offered and point out any gaps and areas for improvement. At the end of the vacation, for this service, Bob receives an NFT certifying him as \textit{friend of Smallville}, which Bob collects along with similar NFTs from other places. At the same time, he receives an anonymous voucher that grants him a hotel discount for the next visit to Smallville.
%attends a concert, where he receives an NFT via a QR code acquired at the venue, which Bob adds to his collection of concert memories. His collection of NFTs becomes a valuable collectible that could be subject to transfer. At the same time, as a participant in the concert, Bob receives an anonymized voucher to be used for public transportation.
\end{example}

The first contribution of this paper is the design of a privacy-aware system enabling PoE generation and management. As an additional contribution, we integrate the PoE with an incentive-based anonymized mechanism providing rewards, thus offering a general-purpose framework for various scenarios,
like those described in Example \ref{ex:intro}.
The proposed, integrated framework decouples the PoE generation and the reward management in order to ensure a high degree of flexibility from the perspective of usability and privacy. From the technical standpoint, we show how to deploy the proposed system with both centralized and distributed solutions. 

The rest of the paper is organized as follows. In Section 2, we briefly discuss the state of the art about Proof of Attendance and incentive mechanisms. In Section 3, we describe the proposed PoE architecture by proposing both a centralized and a distributed architecture. 
In Section 4, we focus on the applicability of our framework to real-world case studies and we comment on a prototype implementation of the PoE management system. In Section 5, some conclusions terminate the paper.

\section{Background and related work}
%
%Stato dell'arte su PoA e meccanismi di engagement e incentivazione.
% Qui metterei POAP ed eventualmente SBT più il discorso WOM per l'incentivazione. Ripesca

The state of the art in PoA relies on technologies such as NFTs created and distributed through smart contracts on a blockchain. One example is given by the Proof of Attendance Protocol (POAP)~\cite{POAP}, which is a proprietary technology providing a distributed application that leverages an Ethereum sidechain called Gnosis. POAP NFTs are managed via web/mobile interfaces and are represented by ERC-721 tokens, including metadata (event, date, time, and location) and an image associated with the event. The claiming process %has an expiration time and 
employs a QR code or a direct link. Each NFT is uniquely identified, transferable, and managed through the POAP wallet.
NFT POAPs have various real-world use cases in the setting of PoA, and they are often used as passes for events, proofs of membership or ownership, event souvenirs and collectibles, digital trophies for achieving milestones~\cite{Hallila_Juffu} (sometimes they are even the subject of real trades).

With a similar purpose, Vitalik Buterin, founder of Ethereum, introduced the concept of soulbound token (SBT)~\cite{SBT_Buterin}, which is a non-transferable NFT associated with a digital identity and representing self-generated certificates.

The Worth One Minute (WOM) platform~\cite{WOM_social} is a centralized system that recognizes social value through the awarding of anonymous digital vouchers with no monetary value, which can be obtained by participating in specific activities considered of social value. Merchants, institutions, and retailers can decide to apply discounts and promotions to holders of these vouchers, thus contributing to the recognition and enhancement of citizens' social contributions.

%A titolo di esempio, un \textbf{WOM Instrument} pu\`{o} rappresentare un sistema di crowd-sensing oppure, nel nostro caso, una soluzione di proof of engament. Esso \`{e} costituito da un server (\textbf{WOM Aggregation Service}) che funge da garante rispetto alle attivit\`{a} dell'utente (il sensing di dati oppure un coinvolgimento in qualche attivit\`{a}), e da un client (\textbf{WOM Collection Tool}) ovvero un software che permette all'utente di raccogliere dati o dimostrare il proprio coinvolgimento. I dettagli di funzionamento di un WOM Instrument vengono lasciati volutamente separati dalla piattaforma WOM (che ne descrive solo l'interfaccia), in modo da poter essere \textbf{implementati in maniera autonoma} da chiunque voglia partecipare ai casi d'uso previsti dal sistema Worth One Minute.
%Anche la piattaforma di gestione dei voucher, la \textbf{WOM Platform}, \`{e} composta da un server (\textbf{WOM Registry}) che genera gli incentivi, e un client (\textbf{WOM Pocket}), ovvero un software che gli utenti utilizzano per collezionare e spendere i voucher presso un qualsiasi commerciante che partecipa al sistema WOM (\textbf{Merchant}). 

The Open Badges technology~\cite{OpenBadges} is an open standard defined by Mozilla Foundation and, nowadays, maintained by the international e-learning consortium 1EdTech. The objective of the project is to facilitate the digital certification of competencies, skills, knowledge, and experience within academic and professional domains. Open Badges aim to have the same legal value as any paper certificate issued by the organizations participating in the consortium. Formally, an open badge is a digital image including JSON metadata and is uniquely linked to information about the recipient, the badge purpose, and details about its acquisition and temporal validity. More recently, Blockcerts~\cite{Bestr} has been proposed as an open standard based on Open Badge 2.0 for the certificate format and on the blockchain technology for storage and management.

\section{System Architecture}

Our PoE framework can be deployed on two different architectures, depending on whether the PoE is managed by a central authority or via a distributed ledger technology. In the following, we describe these two different architectures.

\subsection{Centralized PoE}

The centralized solution to the PoE problem is based on a lightweight handshake that allows a possibly uncertified client to interact with a certified server to transfer data whose ownership can be proven by the client at any time in the future and to any interested third party. Such a data transfer can be either server-client (as in the case of buying a ticket) or client-server (as in cases where the client is motivated to share information with the server, like in crowdsourcing scenarios). The client must always initiate the handshake, while the server must provide its own certificate (as in a one-way Authenticated Key Exchange - AKE~\cite{paar-book}). Depending on the context, mutual authentication could be necessary (as in the case of a certification issue to a legitimate client). At the end of the handshake, the client possesses an ephemeral key pair, which is used later on to show the PoE to third parties (e.g., through zero-knowledge sigma protocols~\cite{MKWK}). At the same time, the server, in turn, associates the public key with the data exchanged during the handshake.

\begin{figure}[t]
    \centering
    \includegraphics[width=1\textwidth]{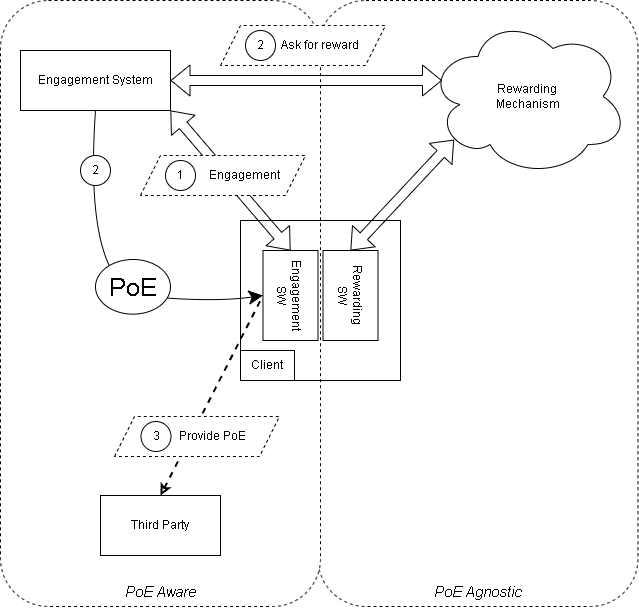}
    \caption{Centralized PoE architecture}
    \label{fig:arch_poe_cent}
\end{figure}

The centralized solution to the PoE management is presented in Figure~\ref{fig:arch_poe_cent}. 
The Client (C) initiates a connection with the Engagement System -- ES (phase 1) by using the Engagement Software -- ESW. Then, ES generates and signs the PoE (phase 2), which is stored locally by the client. If applicable, ES notifies the Rewarding Mechanism -- RM to enable the association of incentives with the engagement. 
Afterwards, the client can exhibit the PoE to any Third Party -- TP an arbitrary number of times (phase 3), and redeem the incentives from the RM in the form of one-time vouchers, by using the Rewarding Software -- RSW. To ensure the confidentiality of the exchanged data and, wherever necessary, the authentication of the involved parties, all the handshakes are based on virtual private networks (VPNs) established via AKE protocols, like, e.g., TLS~\cite{paar-book}. 

\subsubsection{Generating the PoE}

The protocol at the base of the phases 1 and 2 discussed above is defined in Figure \ref{fig:poe-gen}.
%The authentication is mandatory only for the ES.
The session starts with the client sending data to the ES to prove her engagement, see step (2). At this time, C and ES negotiate the degree of data anonymization before building the PoE to balance C's privacy and the accuracy and detail level of the PoE. The result of such a trade-off strictly depends on the application domain and type of PoE (we recall that the PoE may represent a ticket, a badge, a certificate, a proof of ownership, or simply a proof of the engagement of C in some kind of activity). We claim that, in general, the PoE should contain as little information about C as possible, as the PoE will be exposed to third parties (as we will see, the PoE is exposed publicly in the case of a distributed architecture). In many practical scenarios, the PoE may include only essential data together with the integrity proof (e.g., in the form of a hash value) of a detailed data record, which is maintained locally by C and shared with selected third parties on demand. 
After the negotiation, C generates a pair of ephemeral keys, see step (3), a fresh one-time password (which, as we will see, will be used to claim the reward, if any), see step (4), and then sends to ES the public key of the ephemeral pair and the password, see step (5).
At step (6), ES builds and stores locally the PoE, which includes the information agreed with C and the public key chosen by C. At step (7), ES signs the PoE with her secret key $sk_{ES}$, and at step (8), ES sends the PoE and the signature to C.

\begin{figure}[tbh]
\[\begin{array}{lll}
    (1) &  \text{VPN setup} & : C \stackrel{}{\longleftrightarrow} ES  \\[2mm]
    (2) &  C \stackrel{\text{VPN}}{\longleftrightarrow} ES & : (\textit{engagement\_data})  \\[2mm]
    (3) & C & : (\textit{secret\_key}, \textit{public\_key}) \leftarrow Gen_{}()  \\[2mm]
    (4) &  C & : Pwd := \lhd \; \textit{randomly chosen by client} \; \rhd  \\[2mm]
    (5) &  C \stackrel{\text{VPN}}{\longrightarrow} ES & : (\textit{public\_key}, Pwd)  \\[2mm]
    (6) &  ES & : \textit{DB}_{PoE}  \leftarrow \textit{DB}_{PoE} \cup \{ PoE \} \\[2mm]
    (7) &  ES & : \textit{PoE\_sign} \leftarrow Sig(sk_{ES}, PoE)  \\[2mm]
    (8) &  ES \stackrel{\text{VPN}}{\longrightarrow} C & : (\textit{PoE\_sign}, PoE) 
\end{array}\]
\caption{PoE generation protocol}\label{fig:poe-gen}
\end{figure}

\subsubsection{Exhibiting the PoE}

In this phase, Client C uses the PoE through the ESW to prove the engagement, an arbitrary number of times, to any Third Party -- TP. 
A Challenge-Response ID protocol with digital signature \cite{paar-book} is used for the realization of this phase, see Figure \ref{fig:poe-provide}. 

\begin{figure}[tbh]
\[\begin{array}{lll}
    (1) &  C \stackrel{\text{VPN}}{\longrightarrow} TP & : (\textit{PoE\_sign}, PoE)  \\[2mm]
    (2) &  TP & :Ver(pk_{ES}, \textit{PoE\_sign}, PoE) \stackrel{?}{=} \textbf{accept}  \\[2mm]
    (3) &  TP \stackrel{\text{VPN}}{\longrightarrow} C & : ch  \stackrel{R}{\leftarrow} \mathcal{C}  \\[2mm]
    (4) &  C \stackrel{\text{VPN}}{\longrightarrow} TP & : z \leftarrow Sig(\textit{secret\_key}, ch)  \\[2mm]
    (5) &  TP & : Ver(\textit{public\_key}, z, ch) \stackrel{?}{=} \textbf{accept}
\end{array}\]    
\caption{PoE exhibition protocol}\label{fig:poe-provide}
\end{figure}

At step (1), C sends the ES-signed PoE to TP. At step (2), TP verifies the validity of the PoE signature by using the ES public key $pk_{ES}$ and, if accepted, responds to C by sending a random challenge chosen by the challenge domain $\mathcal{C}$, see step (3). At step (4), C uses the secret key corresponding to the PoE to sign the challenge, and then sends it to TP. At step (5), TP uses the public key stored in the PoE to verify the signature provided by C. The PoE is accepted if such a verification succeeds. 

\subsection{Distributed PoE}

The solution discussed above modulates the information exchanged between the parties, thereby ensuring compliance with the privacy requirements dictated by the context or by the client's preferences. Nevertheless, certain constraints may be encountered by the client/server in storing and/or managing the PoE.\footnote{For instance, after issuing the PoE, the ES might not be interested in ensuring its maintenance, including long-term storage or additional services, like transferability, which is a case we have not considered in the centralized scenario.} Therefore, we propose an alternative, distributed solution that integrates established methodologies with DLT-based tools.

In a distributed context, DLT-based solutions to the PoE problem can be realized through NFT technologies and blockchain. Here, we propose two alternatives: in one case, the PoE is stored directly on the blockchain, while in the other case, the PoE is inserted within the metadata of an NFT, which is managed on the blockchain by means of a smart contract.

\begin{figure}[tbh]
    \centering
    \includegraphics[width=1\linewidth]{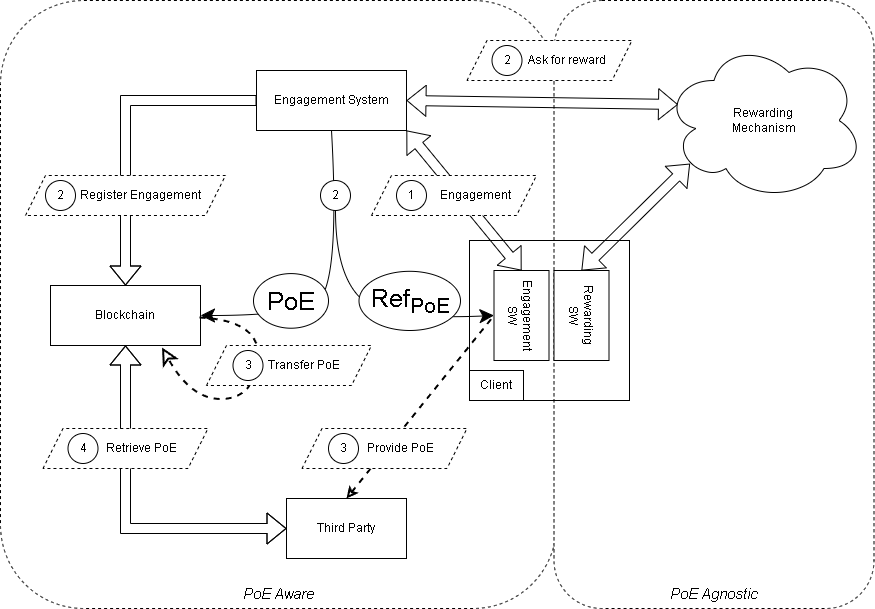}
    \caption{Distributed PoE architecture.}
    \label{fig:arch_poe_dist}
\end{figure}

The general architecture of the distributed solution to the PoE management is presented in Figure~\ref{fig:arch_poe_dist}. Unlike the centralized case, the interaction between C, ES, and TP involves the mediation of a blockchain (BC). Indeed, basically, while the engagement is not different with respect to the centralized scenario (phase 1), the remaining phases related to the PoE management (registration of the PoE by the ES and its sharing with C, exhibition of the PoE by C to any TP, and potential transferability of the PoE) require access to the blockchain, as detailed in the following.

The solution that does not involve the use of NFTs consists of having the server sign a blockchain transaction containing the PoE, once the handshake initiated with the client at the time of engagement is finished. In this way, the server can publish the PoE on the blockchain, which includes the ephemeral public key chosen by the client. 
Hence, the PoE generation protocol is a slight variant of the handshake illustrated in Figure \ref{fig:poe-gen}, where the last three steps are replaced as follows:
\[\begin{array}{lll}
(6) &  ES & : \textit{BC}_{PoE}  \leftarrow \textit{BC}_{PoE} \cup \{ \textit{PoE} \} ; \\
(7) &  ES \stackrel{\text{VPN}}{\longrightarrow} C & : (\mathit{Ref}_{PoE})
\end{array}\]    
Step (6) involves the registration of the transaction on the blockchain that contains the PoE signed by the ES. Finally, at step (7), the ES sends to C the reference to such a transaction.

\begin{figure}[tbh]
\[\begin{array}{lll}
    (1) &  C \stackrel{\text{VPN}}{\longrightarrow} TP & : \mathit{Ref}_{PoE}  \\[2mm]
    (2) & TP & : PoE \stackrel{\text{Ref}}{\longleftarrow} BC \\[2mm]
    (3) &  TP \stackrel{\text{VPN}}{\longrightarrow} C & : ch  \stackrel{R}{\leftarrow} \mathcal{C}  \\[2mm]
    (4) &  C \stackrel{\text{VPN}}{\longrightarrow} TP & : z \leftarrow Sig(\textit{secret\_key}, ch)  \\[2mm]
    (5) &  TP & : Ver(\textit{public\_key}, z, ch) \stackrel{?}{=} \textbf{accept} \\[2mm]
\end{array}\]
    \caption{Blockchain-based PoE exhibition protocol}
    \label{fig:poe-provide-DL}
\end{figure}

The Challenge-Response ID protocol used to exhibit the PoE is a variant of the protocol of Figure \ref{fig:poe-provide} -- see Figure \ref{fig:poe-provide-DL}. The main difference is that in the distributed scenario, the PoE is recovered from the blockchain (see, e.g., ~\cite{9520375}).

%Alternatively, if NFTs are not used, it is sufficient to register on the blockchain a transaction that certifies the assignment of the proof in question to an address, represented by a public key, of which the PoE recipient possesses the secret key counterpart necessary to claim the proof at any time. Hence, the exhibition of the PoE works as usual. Even in this setting, the PoE can be transferred by signing a new transaction, which takes as input the transaction of registration of the PoE, by using the private key associated with that proof. 

The NFT-based solution is inspired by the buying and selling system of NFTs of POAP. Once the blockchain technology to be used is chosen and users are equipped with the digital wallets associated with their blockchain account, a smart contract is employed to manage the creation, transfer, and/or ownership verification of the NFT associated with the PoE. The engagement data are exchanged during a client-server handshake and are stored outside the blockchain, but can be retrieved via the associated NFT. The data integrity is guaranteed by the immutability of the NFT via a cryptographic hashing mechanism. 
After the engagement, the ES proceeds to register the proof thus generated within the BC and to return to the Client (via the ESW) the reference to the BC where to retrieve the PoE (which is no longer stored by the Client as in the centralized case). 
The secret key associated with a specific user's digital wallet is used to claim ownership of the proof to anyone with read access to the blockchain. The association of the public key (related to a user's wallet) with the engagement data is ensured by the corresponding NFT, by using the smart contract that manages the NFTs. To this aim, it is sufficient for the Client to send to the TP the reference to the proof that must be shown. 
%In this architecture, the Client can also transfer the PoE paternity to another user by using the smart contract managing the NFT. The rest of the architecture remains the same as that shown in the centralized case.

\subsubsection{PoE transferability}

As in the classical NFT marketplace, we assume that, depending on the application domain, the ownership of certain PoEs could be transferred, along with the ability to subsequently claim the ownership.
%Even if the PoE were to be transferred to a user other than the original recipient of the PoE, the ability of that user to provide the PoE (which he now possesses) to any TP must be guaranteed. So, in case of transfer, a problem arises: if the PoE has been transferred at least once, then the public key associated with the PoE, generated by the current owner of the PoE, is different from the PoE public key generated at the time of engagement. Therefore, it is necessary to replace the pair of PoE ephemeral keys. In the following, we show the transfer protocol, without/with the involvement of NFTs.

The solution without NFTs is a general-purpose one inspired by the Bitcoin protocol~\cite{Princeton_Bitcoin} and can be adapted to any blockchain. The protocol is shown in Figure \ref{fig:poe-trans}, where C is the current owner of the PoE, R is the recipient to whom the ownership of the PoE is transferred, and $T$ is the transaction establishing the current ownership of the PoE.
At step (1), R chooses the pair of ephemeral keys associated with the transfer, while at step (2), the public key of such a pair is sent to C. Then, at step (3), C builds the transfer transaction stating that the PoE is now associated with a new public key. Such a transaction is signed by C with the ephemeral secret key associated with $T$, and then written to the blockchain, see step (4).

\begin{figure}[tbh]
\[\begin{array}{lll}
    (1) &  R & : (pk_{rec}, sk_{rec}) \leftarrow Gen()  \\[2mm]
    (2) & R \stackrel{\text{VPN}}{\longrightarrow} C & : (pk_{rec}) \\[2mm]
    (3) &  C & : T' := \{ pk_{rec} \text{ is the new owner of the } \textit{PoE} \text{ at transaction T} \}  \\[2mm]
    (4) &  C & : BC \leftarrow Sig(\textit{secret\_key}, T')
\end{array}\]
    \caption{PoE transfer protocol without NFT}
    \label{fig:poe-trans}
\end{figure}

Whenever the PoE is stored as an NFT, the machinery managing the transfer is left to a smart contract. 
Starting from the interface of the ERC-721 standard~\cite{Erc-721,10.1145/3366611.3368142}, it is possible to implement a smart contract that manages the NFT associated with a PoE. In this solution, the paternity of the PoE can be made transferable by exchanging the corresponding NFT between the wallets of any two users (see, e.g., \cite{10152551}).
%This stage does not involve intermediaries outside the blockchain.

\subsection{PoE with rewards}

\begin{figure}[tbh]
    \centering
    \includegraphics[width=1\linewidth]{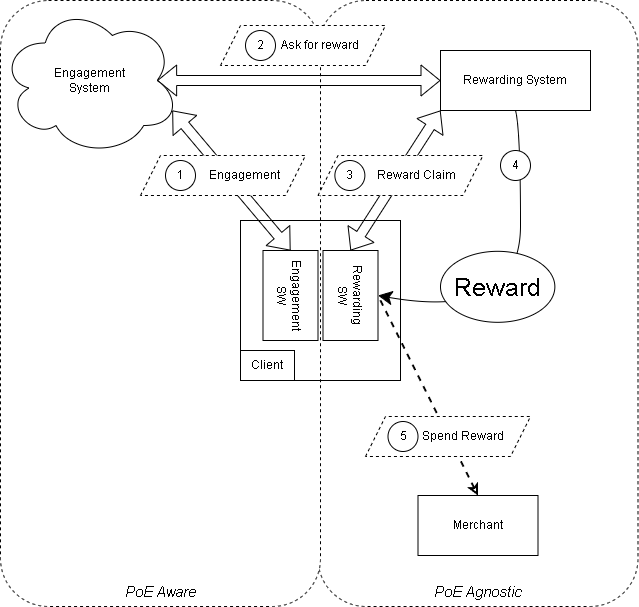}
    \caption{Reward mechanism architecture.}
    \label{fig:arch_poe_reward}
\end{figure}

The system we propose envisages the possibility of creating a PoE associated with incentives in the form of rewards. The Rewarding Mechanism -- RM used in both PoE architectures (centralised and distributed) is shown in Figure \ref{fig:arch_poe_reward}.

In this system, after the handshake between C and ES certifying some kind of engagement (phase 1), another handshake is activated
between ES and the Rewarding System -- RS (phase 2). In response,  RS generates the reward corresponding to the ES request. The client then proceeds to redeem the reward through a Reward Claim (phase 3) by using the Rewarding Software -- RSW. The reward is then delivered (phase 4) and subsequently spent (phase 5) by the client in any interaction
with merchants accepting the corresponding type of incentive. The RS guarantees that a reward cannot be used more than once and that no more than one reward can be generated for the same PoE. 

%%% PRIVACY
%Il \textbf{livello di privacy} garantito da questo sistema di PoE corredato da incentivi viene \textbf{negoziato} tra Client ed Engagement System al momento dell'engagement. Essi, una volta terminata l'azione da certificare ed incentivare, si mettono d'accordo sui dati di engagement da inserire all'interno della PoE (che potrebbero divenire pubblici nel caso di PoE distribuita) in modo da \textbf{tutelare la privacy del Client} che rimane in ogni caso \textbf{anonimo}, in quanto, nella PoE non vengono inseriti dati che possano identificarlo. Di contro, se richiesto dal caso d'uso, \`{e} possibile (ma non obbligatorio) rendere la PoE \textbf{indissolubilmente legata} al suo possessore (\textbf{riducendo} il livello di privacy garantito al Client) inserendo nella PoE degli specifici identificativi relativi all'utente destinatario della prova. 

Independent of the specific use case and the PoE architecture, it is worth observing that RS and reward are designed to be agnostic of the PoE, meaning that they do not manage or include any information about the client or the details of the PoE. During phase 2, ES guarantees that a given PoE, which is not shown to the RS, is eligible to issue a certain type of reward. Hence, the reward is completely anonymous.

%Qualsiasi sia il caso d'uso, Rewarding System, Rewarding Software e Merchant sono stati progettati per essere \textbf{PoE Agnostic}, cio\`{e}, hanno bisogno di conoscere solamente una \textbf{piccola porzione} dei dati di engagement (quelli necessari per creare il Reward corrispondente) mentre il resto delle informazioni riguardanti la PoE rimangono \textbf{private} (\textbf{inclusa}, se presente, \textbf{l'identit\`{a}} di chi ha effettuato l'engagement). Questo \`{e} utile nel caso in cui il Rewarding System sia in grado di preservare la \textbf{privacy degli utenti (che rimangono anonimi)} e in particolare possa creare i Reward senza conoscere tutti i dettagli riguardanti la PoE ma solo quelli forniti dall'ES. Questi ultimi dati vengono \textbf{precedentemente concordati} con il Client nella fase di Engagement e sono successivamente \textbf{anonimizzati} ad opera dell'ES. Di fatto, anche qualora le informazioni relative ad una PoE divenissero pubbliche, \textbf{nessuno di essi} (nemmeno il RS) sarebbe in grado di \textit{associare} ad un Reward la PoE da cui esso deriva.

% On the other hand, the ES system is said to be \textbf{PoE Aware}, in quanto, pur essendo a conoscenza di tutti i dati inseriti all'interno della PoE, non possono in alcun modo \textit{associarli} al corrispondente \textbf{Reward}, che pu\`{o} essere utilizzato in maniera \textbf{anonima}. 

%We now present the details of the handshake protocol for generating and redeeming the reward, which is not affected by the specific centralized/distributed architecture used for the PoE management.

\subsubsection{Generating and Redeeming the Reward}
The protocol at the base of the phases 2-4 discussed above is defined in Figure \ref{fig:rew-gen}.

\begin{figure}[tbh]
\[\begin{array}{lll}
    (1) &  ES & : \textit{filtered\_data} \leftarrow \textbf{Anonymize}(\textit{engagement\_data})  \\[2mm]
    (2) &  ES \stackrel{\text{VPN}}{\longrightarrow} RS & : (\textit{filtered\_data}, H(Pwd))  \\[2mm]
    (3) &  RS \stackrel{\text{VPN}}{\longrightarrow} ES & : (OTC, H(Pwd))  \\[2mm]
    (4) &  ES \stackrel{\text{VPN}}{\longrightarrow} C & : (OTC, H(Pwd))  \\[2mm]
    (5) &  C \stackrel{\text{VPN}}{\longrightarrow} RS & : (OTC, Pwd)  \\[2mm]
    (6) &  RS \stackrel{\text{VPN}}{\longrightarrow} C & : (Reward)  \\[2mm]
\end{array}\]
    \caption{Reward generation protocol}
    \label{fig:rew-gen}
\end{figure}

At step (1), ES extracts from the PoE engagement data the minimum amount of information needed to issue a reward request to the RS (e.g., the type of eligible request, which allows the RS to specify in which context a reward can be spent).
Then, the handshake between ES and RS is based on an AKE protocol with mutual authentication. At step (2), the ES transmits the digest $H(Pwd)$ (computed using the hash function $H$) of the one-time password \textit{Pwd} associated with the PoE by the client (see Figure~\ref{fig:poe-gen}), together with the filtered data. This will allow the client to redeem the reward later on. Then, the RS issues the reward and sends to the ES the One Time Code (OTC) for accessing it, see step (3). At step (4), the ES forwards such information to the client within the same handshake session related to the PoE generation, e.g., in step (8) of Figure~\ref{fig:poe-gen} (together with the signed PoE). In steps (3) and (4), the hashed password acts as a nonce associated with the OTC. 
Afterwards (and by the expiration time associated with the reward), the client employs the OTC to access the reward and redeem it by exhibiting the corresponding $Pwd$ as proof of claim, as in classical password-based identification protocols \cite{paar-book}, see steps (5) and (6). Note that to transmit the reward, the digest of the received password must correspond to the locally stored value associated with the related OTC.
%This is done in the setting of a VPN instantiated between C and RS, through which the RS transmits the reward to the client -- see step (6) -- if the digest of the received password corresponds to the locally stored value associated with the related OTC.

The protocol presented above is inspired by the WOM platform, to which our system can be integrated. For the presentation of the details of the one-time spending process of the WOM vouchers, the interested reader can refer to~\cite{Klopfenstein2019}.

\section{Case studies}

In this section, we describe a prototype of the proposed system that has been implemented for a specific case study, and we discuss potential integration with other platforms and in several real-world application domains.

\subsection{Simulation}

To simulate the behavior of the proposed system, we have implemented a prototype system managing the PoE in the case of attendance at seminars, which is a case study already under development at the University of Urbino for the assignment of WOM vouchers.

The software was developed using the programming language Dart and the Google Flutter framework. It includes three modules implementing the centralized solution to the PoE and deployed as three web applications representing Client\footnote{\url{https://github.com/MatteoMarcoM/poe_client}}, Engagement System\footnote{\url{https://github.com/MatteoMarcoM/poe_es}} and Third Party\footnote{\url{https://github.com/MatteoMarcoM/poe_tp}}. The Client module allows the user to collect and manage PoEs resulting from the interaction with the Engagement System. At the same time, the module allows the user to exhibit the PoE to the Third Party, which verifies its validity. The three web apps interact in a virtual LAN through a server\footnote{\url{https://github.com/MatteoMarcoM/web_socket_dart}} that follows the WebSocket protocol.
The PoE includes student ID, name, surname, email, GPS data, and timestamp. Once approved by the Engagement System web app, the PoE is generated, signed, and sent to the Client. By following the corresponding protocol, the Client can choose a PoE from the wallet managed by the web app and exhibit it to the Third Party, which performs the validation check.

The PoE format is based on JSON (JavaScript Object Notation)~\cite{JSON_Org} and the standard ECMA-404 -- chosen for its flexibility and interoperability -- and includes the following information (see the example illustrated in Figure~\ref{fig:poe-json}):

\begin{itemize}
\item The public key of the ephemeral key pair associated with the PoE generation, which is implemented using the RSA scheme;
\item Spatial (e.g., GPS coordinates) and temporal (e.g., timestamp of the engagement) information (if eligible);
\item Engagement data as negotiated between the Client and the Engagement System;
\item Any other structured data (e.g., images) following the JSON standard;
\item The integrity proof (expressed as the digest deriving from the application of a one-way hash function, like, e.g., SHA-256) of additional data that are not stored directly within the PoE but are maintained locally by the Client web app;
\item The \textit{transferability} flag stating whether the PoE can be transferred to another client;
\item The expiration date of the proof (if eligible).
\end{itemize}

\begin{figure}[tbh]
    \centering
{\scriptsize \begin{lstlisting}
JSON = {
    "proof_type" : "PoE",
    "transferable" : true,
    "public_key" : {
        "algorithm" : "SHA-256/RSA",
        "verification_key" : "AAAAB3NzaC1yc2EAAAABJQAAAQB/
                              OWmcKS0A8 ... x836Sj/6LcjQ8n" 
    },
    "timestamp" : {
        "time_format" : "UTC",
        "time" : "2005-10-30 T 10:45"
    },
    "gps" : {"lat" : -34, "lng" : 151, "alt" : 1200},
    "engagement_data" : {
        "encoding" : "base64",
        "data" : "TG9yZW0gaXBzdW0gZG9sb3Igc2l0IGFtZX
                  QsIGNvbnNlY3RldHVyIGFkaXBpc2NpIGVs
                  BpbmNpZHVudCB1d ... YWJvcmUgZXQgZG"
    },
    "sensitive_data" : {
        "data_1" : "<HASH(value_1)>",
        "data_2" : "<HASH(value_2)>",
        ...
        "data_n" : "<HASH(value_n)>"
    },
    "other_data" : {
        "expiration_date" : {
            "date_format" : "UTC",
            "date" : "2035-10-30 T 10:45"
        }
        ...
    }
}
\end{lstlisting}}
\caption{Example of a PoE in JSON format.}
\label{fig:poe-json}
\end{figure}

\subsection{Application domains}

The context that provided us with the motivational basis for proposing a general-purpose framework for PoE management is the project \textit{Casa delle Tecnologie Emergenti di Pesaro} (CTE Square)~\cite{CTE_Square}, which was conceived as an advanced center for technological innovation and skills transfer, where the application of emerging technologies in the areas of Culture, Tourism and Engagement can be tested. In this setting, the PoE is conceived as a framework
serving to cover the temporal and spatial distance between the action to be recognized, the recognition of the corresponding validity, and the application of incentives, by achieving an adequate trade-off between fairness, usability, and privacy protection. The technology identified for realizing the reward-based incentive mechanism is the WOM Platform~\cite{Klopfenstein2019,WOM_remedi,WOM_social}. Two case studies that emerged in the CTE Square project are illustrated as follows.

\subsubsection{University exams certification}
In a way inspired by the Open Badge solution, a PoE is provided to the student upon passing an exam. In this case, the ES is implemented within the University platform that manages the students' careers. The PoE is flexible enough to include customized information detailing competences acquired or projects developed to pass the exam, which represents very specific data that is usually not maintained by the University. Such a PoE can be integrated into the student's CV and diploma supplement. In addition, as a social incentive, for each exam that is passed, a reward is assigned in the form of anonymous education vouchers that can be spent to obtain discounts in partner shops and stores, such as bookstores and cafes. The same approach is applied to certify and incentivize participation in other extracurricular events and activities (seminars, soft skills courses, job placement workshops, and many more). In this case study, the proposed solution is centralized, where the ES acts as the central certifying authority, and the PoE is not transferable.

\subsubsection{Tourism office initiatives} 
In a way inspired by the POAP NFT solution, a PoE is provided to the users contributing actively to improving the quality of the local tourist and cultural receptivity (e.g., by visiting tourist sites, providing feedback, and supporting events). The specificity of the PoE is that it is anonymous and may include exclusive and individual contents representing collectibles whose ownership could be transferred. In addition, anonymous vouchers are emitted that entitle the holder to discounted access to exhibitions and other events.
In this case study, the proposed solution is distributed and blockchain-based, and the PoE is transferable.

%\subsubsection{PoE added value}

\section{Conclusions}

In this work, we analyzed and designed flexible, general-purpose, and privacy-aware solutions to the Proof-of-Engagement problem, integrating also incentive aspects and combining advanced technologies, ranging from Distributed Ledger Technologies to Non-Fungible Tokens. 
The main strengths of adopting our proposed PoE solutions can be summarized as follows:
\begin{itemize}
    \item Flexibility: PoE solutions allow a wide range
of data types to be stored within the proof, making the system adaptable to
multiple use cases and application scenarios.
\item Ease of integration: the implementation of such solutions is facilitated by
standardized and well-established technologies that allow minimizing
implementation time, even in contexts that involve integration with other platforms, like, e.g., WOM.
\item Efficiency and privacy protection: PoE technologies ensure high
operational efficiency, while still maintaining an adequate level of privacy
guaranteed to users, thanks to the use of tools designed specifically
for this purpose.
\end{itemize}

The centralized architecture offers simplicity and rapid deployment in different scenarios, which is ideal in controlled enterprise and corporate environments. The distributed architecture, on the other hand, provides transparency, data immutability, and scalability, which are features particularly required in public and collaborative settings. The integration of the PoE with incentive mechanisms has also shown considerable potential in scenarios requiring interaction and enhancement of human behaviors (like, e.g., in crowdsourcing), with positive implications for large-scale applications.

The proposed framework is currently under development for the deployment of digital services supported by the municipality of the city of Pesaro and by the University of Urbino, Italy. These services aim to promote tourism, culture, and engagement in activities recognised for their high social value.
As future work, it is worth completing the deployment of the framework and its full integration with the WOM platform, in order to test its efficacy and efficiency in large-scale scenarios and various application domains, including environmental sustainability initiatives, social awareness campaigns, and global educational programs. 
	
\bigskip
{
\large \textbf{Acknowledgements}} \\
This work is partially funded by MIMIT, under FSC project “Pesaro CTE SQUARE”, CUP D74J22000930008.

\bibliographystyle{splncs04}
\bibliography{bibliography}
\end{document}